\documentstyle[aps,prb,multicol,epsf]{revtex}
\begin{document}
\draft

\title{Enlarged Symmetry and Coherence in Arrays of Quantum Dots}
\author{A. V. Onufriev and J. B. Marston}
\address{Department of Physics, Brown University, Providence, RI 02912-1843}
\date{\today}
\bigskip
\maketitle

\begin{abstract}
Enlarged symmetry characterized by the group $SU(4)$ can be realized 
in isolated semiconducting quantum dots.  A Hubbard 
model then describes a pillar array of coupled dots and at half-filling
the system can be mapped onto an
$SU(4)$ spin chain.  The physics of these new structures is rich as
novel phases are attainable.  The spins spontaneously dimerize and   
this state is robust to perturbations which break $SU(4)$ symmetry.
We propose ways to experimentally verify the existence of the dimerized phase. 
\end{abstract}

\bigskip

\section{Introduction}
\label{intro}

Quantum dot arrays are a new arena for the study of strongly correlated 
electrons and the persistence of quantum coherence.  Physical properties of
a single semiconducting dot as well as tunneling between dots can be controlled over a wide
range -- a luxury not available to us in ordinary condensed materials.  Recent
advances in nanofabrication techniques offer the possibility of constructing 
artificial structures so small that the electronic level spacing is
comparable to the charging energy.  As a consequence, these structures
can exhibit enlarged continuous 
symmetries not normally found in nature. In this paper   
we determine conditions under which a pillar of coupled semiconducting
quantum dots realizes the group $SU(4)$ as a good symmetry and show that
the $SU(4)$ spins spontaneously dimerize --  
a phase of matter that would be difficult to attain with the smaller 
$SU(2)$ symmetry of electrons in generic quantum dots.  

Continuous symmetries are ubiquitous in physics.  Rotational invariance
characterized by the group $O(3)$
permits the classification of atomic orbitals via integer
angular momentum quantum numbers\cite{Baym}.
Spinning particles, such as electrons,
are described by representations of the group $SU(2)$.
Approximate $SU(3)$ isospin symmetry of hadrons has its origin
in the light masses of the up, down and
strange flavors of quarks\cite{Itzykson}.  Unlike
the case of electrons for which $SU(2)$ symmetry is exact, quarks
can be described by  $SU(3)$  only approximately since the masses of 
the quarks are not exactly equal.
Nevertheless, the approximate $SU(3)$ symmetry is useful for
classifying hadrons.  We show how an approximate 
$SU(4)$ symmetry can be realized in quantum dot structures, and we exploit 
its properties to describe the novel dimerized 
phase that should emerge in these structures under certain conditions.   
 
Consider a potential well with $N$ degenerate 
eigenstates.  Taking electron spin into account   
there are a total of $2 \times N$ degenerate states, and if 
all of these states are equivalent, in a sense made precise below,
we can think of them as realizing the fundamental representation of
the $SU(2N)$ group.  In other words, electrons placed in the shell can be 
considered as having $2N$ different, but equivalent, flavors instead of 
just the ordinary two flavors of spin up and down.  It is important to note that
$SU(2N)$ symmetry is {\it not} equivalent, in general, to the higher-spin
representations of the usual $SU(2)$ group familiar from the quantum theory of 
angular momentum.  Rather, for $N > 1$, $SU(2N)$ is a different and larger
symmetry.

Ordinary atomic orbitals might seem like 
a good candidate but, for real atoms, the enlarged symmetry is
broken down to the usual $SU(2)$ symmetry by electron-electron interactions 
which lift the degeneracy.  
However, as Stafford and Das Sarma noticed\cite{DasSarma}, semiconducting
quantum dots offer the possibility of realizing enlarged symmetries.
Quantum dots can be thought of as artificial atoms\cite{Kastner,Ashoori}
with tunable parameters.  To be precise, the electron mass is replaced by the 
smaller band mass $m_e \rightarrow m_b$,  
and the Bohr radius $a_B = \hbar^2 / (m_e e^2)$ is replaced by $a_B^* = 
\varepsilon~ (m_e / m_b)~ a_B$.  In $GaAs$, $m_b \approx 0.067 m_e$, the 
dielectric constant $\varepsilon \approx 13$, and $a_B^* \approx 100 \AA$ which
is two orders of magnitude larger than its fundamental value.  
Electrons in a quantum dot are confined in a non-singular potential often
described\cite{Goldman-1} as a short square well in the $z$-direction 
and a simple parabola in the $x-y$ plane, 
$V(x, y) = \frac{1}{2}~ m_b~ {\omega_0}^2 (x^2 + y^2)$, 
though our results do not
depend on the detailed form of the potential as long as it has cylindrical
symmetry.  For the lowest mode in the z-direction,
the resulting harmonic oscillator eigenenergies are $E_{n, l_z} = \hbar 
\omega_0 (2n + |l_z| + 1)$ where $n$ and $l_z$ are respectively the 
radial and angular momentum quantum numbers.   

\section{Constructing and Coupling $SU(2N)$ Quantum Dots}
\label{section2}

\begin{figure}
\begin{multicols}{2}
\epsfxsize = 7.5cm \epsfbox{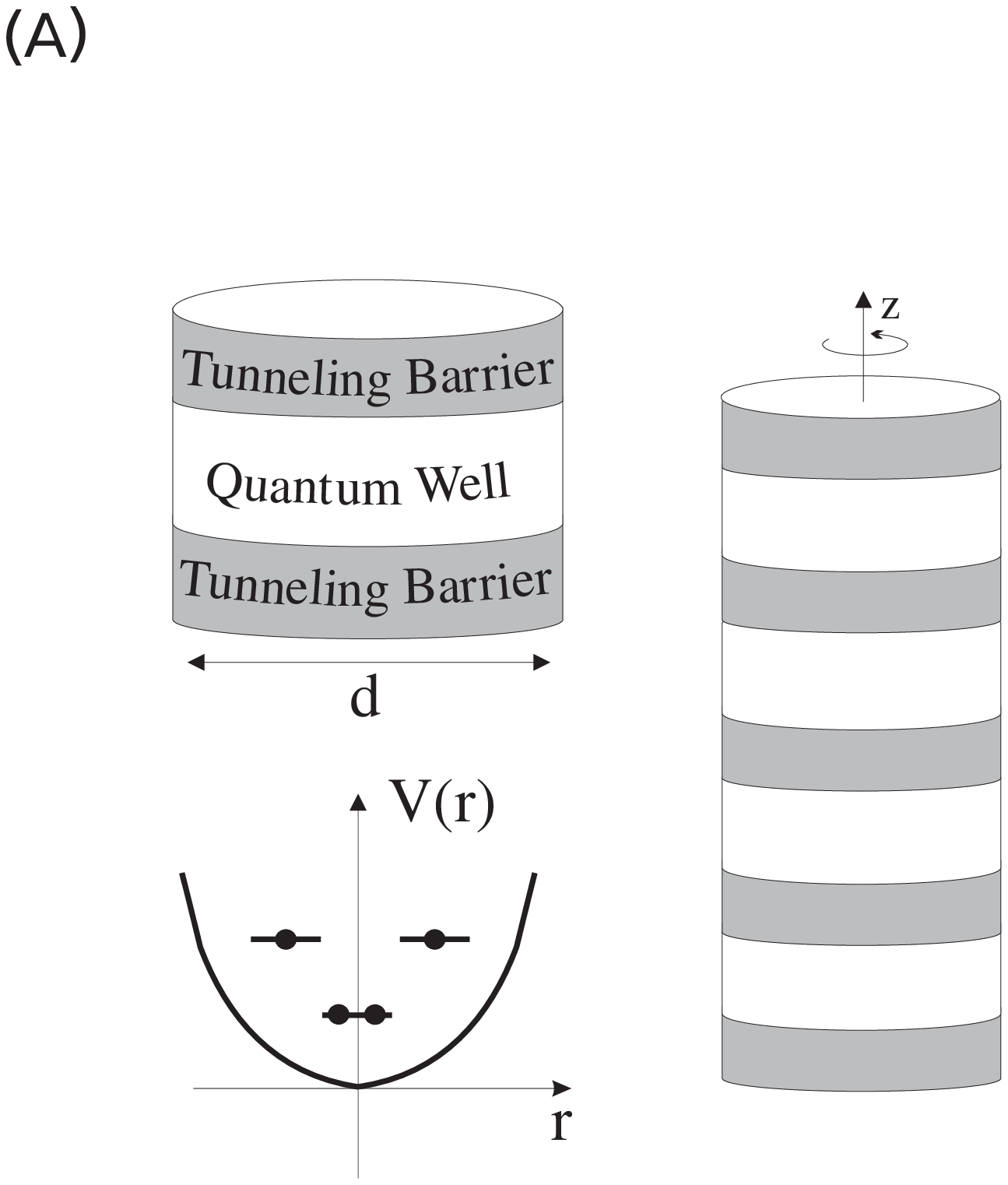}
\epsfxsize = 7.5cm \epsfbox{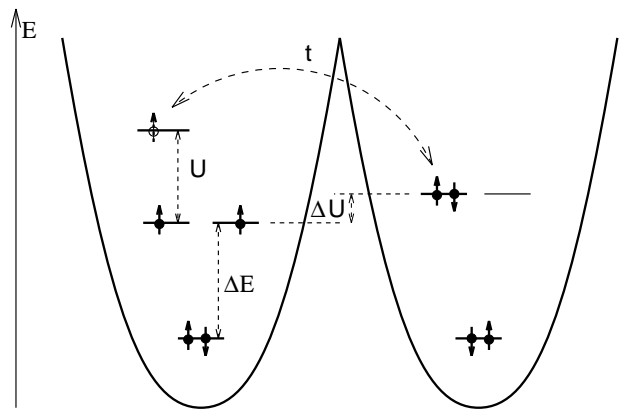}
\end{multicols}
\caption{
(A) Individual quantum dot and proposed pillar array of quantum dots
with approximate $SU(4)$
symmetry. By adjusting the bias, the lowest $s$-level in each dot is filled
completely and the first $p$-level is half-filled with two electrons.
(B) Energy diagram of two adjacent dots from the array. $\Delta U$
is the energy difference between the highest state of total orbital
angular momentum
$|L_z| = 2$ (shown in the right) dot and the lowest state with $L_z = 0$
(depicted in the left dot).  An electron in a $p$-level can temporary hop
into an empty level on an adjacent dot and then hop back.  This virtual
exchange process lowers its energy by of order $J = 4t^2/U$.
If $U \gg \Delta U$ and $J \gg \Delta U$,
all 6 configurations on each dot participate
equally in the exchange and the array realizes approximate $SU(4)$ symmetry.}
\label{fig1}
\end{figure}

We propose a one-dimensional array of rotationally symmetric 
III-V semiconducting quantum dots arranged in 
a pillar\cite{Pawel} and show how approximate 
$SU(4)$ symmetry can be realized in the structure.  
Given sufficient control over dot diameters and 
gate positions and biases\cite{Ashoori} the lowest $s$-level ($n = l_z = 0$)
of each dot can be completely filled (with two electrons), 
and the next higher, four-fold degenerate $p$-level with $n = 0$ and
$l_z = \pm 1$, can be half-filled with two valence electrons\cite{Kouwenhoven},
as shown in  Fig. \ref{fig1}. Apart from configuration splitting
(discussed below), the two $p$-electrons realize a self-conjugate 
(particle-hole symmetric) representation of $SU(4)$.
The dimension of this representation is $6$ which corresponds to the six
distinct ways the two electrons can be placed in the four available 
$p$-states\cite{fundamental}. 
Once electron tunneling between the dots is turned on, there 
are four energy scales in the problem:
the gross level spacing $\Delta E = \hbar \omega_0$ in each well, the on-site 
Coulomb repulsion energy $U$ 
which represents the energy cost to add an additional electron to the dot, 
the energy splitting between the six different $p$-level configurations  
$\Delta U$, and the tight-binding electron hopping amplitude $t > 0$ 
between states in adjacent wells, see Fig. \ref{fig1} (B).  

The advantage of proposed pillar array Fig. \ref{fig1} (A)
is clear:  Conservation of the 
electron's orbital angular momentum around the $z$-axis, a consequence of the 
cylindrical symmetry of the confining potential, guarantees that 
transitions between different angular momentum states in adjacent dots,
which would break the flavor symmetry, 
are forbidden.  The crucial conditions are that the energy gain $J$ due
to electron exchange between the dots greatly exceed $\Delta U$, and also
that the four flavors of electrons participate in the exchange on an 
equal footing.  To second order in perturbation theory\cite{Burkard} 
$J = {4t^2/U}$ and thus we require:
\begin{equation}
\frac{4t^2}{U} \gg \Delta U ~~{\rm and}~~ U \gg \Delta U\ .
\label{condition-2}
\end{equation}
Another inequality ensures that only the $p$-electrons play
an active role in the low energy physics:
\begin{equation}
\Delta E \gg t\ .
\label{condition-1}
\end{equation}
To estimate the size of $\Delta U$ for the $p$-states in a quantum dot 
we first note that spin-orbit coupling is negligible\cite{Note1}.
Thus, $\Delta U$ is due almost entirely to the 
dependence of the electron-electron Coulomb interaction on the 
shell configuration as described by Hund's rules\cite{Baym2}, 
which have been shown both experimentally\cite{Kouwenhoven} and 
theoretically\cite{Koskinen,Steffens}  
to be directly applicable to semiconducting dots.  
The six configurations break up as: $6 \rightarrow 3 \oplus 1 \oplus 2$. 
The triply degenerate
state of total orbital angular momentum $L_z = 0$ and total spin $S = 1$ 
is lowest in energy, the intermediate non-degenerate state has 
$L_z = 0$ and $S = 0$, and the two-fold degenerate highest level has 
$|L_z| = 2$ and $S = 0$.  As both $\Delta U$ and $U$ scale as $d^{-1}$ with 
dot size $d$, we may introduce $\Gamma \equiv \Delta U / U$, where 
$\Gamma$ depends only the shape of the dot and the confining potential.  
Lowest order direct and exchange interaction integrals allow us to  
estimate that $\Gamma$ ranges from $0.5$ for thin, quasi-two-dimensional 
dots to $\Gamma = 0.2$ for thick dots. 
Because the Coulomb interaction is long-ranged, these numbers are nearly 
independent of the 
confining potential; indeed, $\Gamma \approx 0.2$ also holds for real 
(nearly spherical) atoms.   
For an electron in a potential well of characteristic size $d$ we have: 
$\Delta E \approx h^2 / (m_b d^2)$,  
$U \approx e^2 / (\varepsilon d)$, and thus ${{\Delta E}/U} \approx a_B^*/d$.
Symmetry breaking effects due to the electron-electron interaction are
therefore minimal in sufficiently small dots.
To satisfy Eqs. (\ref{condition-2}) and (\ref{condition-1}) with
$\Gamma = 0.2$, simple algebra shows that we require 
$(\Delta E / U)^2 \gg 1$ which is in fact satisfied by small dots.
For example, $InAs/GaAs$ ($a^*_B \approx 300 \AA$) quantum dots 
have been made\cite{Fricke} which have $d \approx 200 \AA$, 
$U \approx 18$ meV, $\Delta E \approx 50$ meV, and
$(\Delta E / U)^2 \approx 8$.  Adjusting the array spacing
and the thickness of the insulating barriers 
the hopping amplitude may be increased\cite{Livermore} to
$t = 0.2~ \Delta E$. 
It then follows that $J \approx 20$ meV. For dots which are not
too thin, $\Gamma = 0.2$, $\Delta U \approx 4$ meV,
and the crucial inequalities Eqs. (\ref{condition-2})
are satisfied.  In contrast to these artificial atoms,  
there are no real atoms for which both 
inequalities Eq. (\ref{condition-2}) and Eqs. (\ref{condition-1}) hold 
because $\Delta E \approx U$ and $t \ll U$. 
A typical example is a copper-oxide antiferromagnet\cite{LesHouches91} with
$J \approx 0.13$ eV, $U \approx 10.5$ eV and hence $\Delta U \gg J$.    

\section{Arrays of $SU(4)$ Quantum Dots}
\label{section3}

Provided that the conditions outlined above are met, the 
pillar array of quantum dots  may be described by an
$SU(4)$ invariant Hubbard model. We retain only nearest-neighbor 
hopping and on-site Coulomb repulsion, and assume that no spin-flip or
orbital-flip processes occur.  Interdot Coulomb repulsion is not expected to
change our results qualitatively.  We use one Greek index $\alpha = 1,...,4$ 
to label all four flavors of $p$-states\cite{DasSarma}: 
$| l_z = 1, s_z = +1/2 \rangle \rightarrow | \alpha = 1 \rangle$, 
$| 1, \downarrow \rangle \rightarrow | 2 \rangle$, 
$| -1, \uparrow \rangle \rightarrow | 3 \rangle$, and 
$| -1, \downarrow \rangle \rightarrow | 4 \rangle$. 
At half-filling the Hubbard Hamiltonian, for an open chain of length $L$ sites,
may be written: 
\begin{equation}
H = \sum_{i=1}^{L-1} \bigg{\{}~ t~ (c^{\dagger \alpha}_i
c_{i+1, \alpha} + H.c.) + U [n(i) - 2]^2 \bigg{\}}\ ; 
\label{su4Ham}   
\end{equation}
here repeated raised and lowered Greek indices are summed over, 
index $i$ labels the dots, $c^{\dagger \alpha}_i$ is the 
creation operator for an electron in state $\alpha$, and
$n(i) \equiv c^{\dagger \alpha}_i c_{i, \alpha}$ is the total
electron number operator at site $i$. 
Both the hopping and interaction terms in the Hamiltonian Eq. (\ref{su4Ham}) 
are explicitly $SU(4)$ invariant as can be
easily checked by applying a unitary transformation, $c^{\dagger \alpha}_i
\rightarrow U^\alpha_\beta c^{\dagger \beta}_i$, 
with $U^{\dagger} U = 1$, which leaves Eq. (\ref{su4Ham}) unchanged.

\begin{figure}
\center
\epsfxsize = 6.5cm \epsfbox{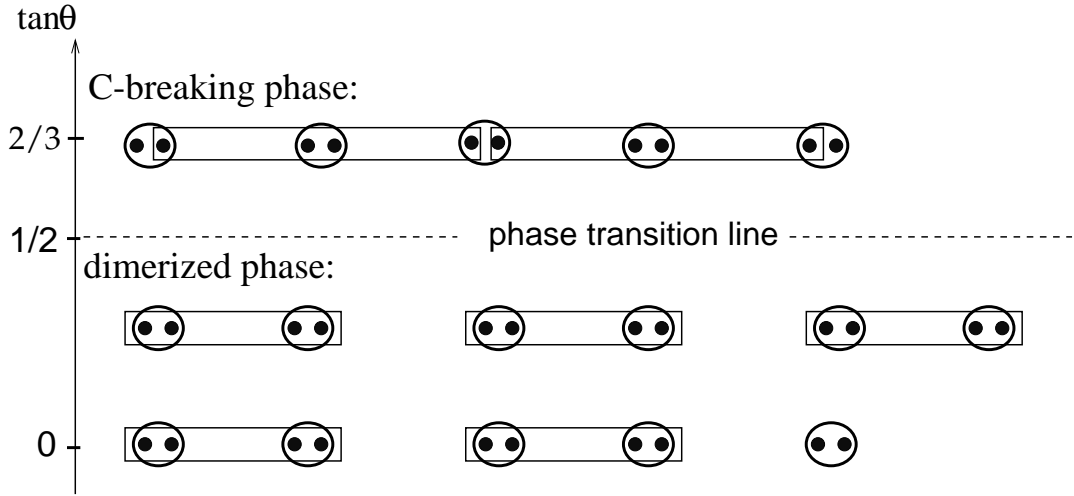}
\caption{
Antiferromagnetic part of the $SU(4)$ phase diagram
for the isotropic, nearest-neighbor $SU(4)$ spin chain.
The two valence $p$-electrons on each quantum dot are depicted
by filled circles.  $SU(4)$ singlet bonds encapsulate four electrons
and are depicted as rectangles.
Chains with an odd number of artificial atoms have free spins at the
chain ends in both the dimerized and the charge-conjugation (C-breaking) phases.
For an even number of atoms, however, the dimerized phase has no free spins.}
\label{fig2}
\end{figure}

At half-filling, the low energy physics of the system is governed by the 
$SU(4)$ spin degrees of freedom as creation of a charge excitation is 
energetically unfavorable.  A weak-coupling renormalization-group (RG)
calculation shows that umklapp scattering processes (which in the 
$SU(4)$ case carry both charge and spin) drive the Hubbard model into
a Mott-Hubbard insulating phase with gaps in both sectors\cite{Marston2}.  
In the strong-coupling limit of ${|t|} / U \ll 1$, again there is a 
charge gap and perturbation theory maps directly
the Hubbard Hamiltonian Eq. (\ref{su4Ham}) onto an insulating 
quantum antiferromagnetic
Heisenberg spin chain. To $O(t^4/U^3)$ the effective Hamiltonian is:
\begin{equation}
\label{spinHam}
H_{SU(4)} = {{J}\over{2}}~ \sum_{i=1}^{L-1}~ \bigg{\{} 
\cos(\theta)~ Tr \{ S(i) S(i+1) \}
+ {1\over 4} \sin(\theta)~ [Tr \{ S(i) S(i+1) \}]^2 \bigg{\}}
\end{equation}  
plus next-nearest-neighbor terms.  Here
$S^{\alpha}_{\beta}(i) = c^{\dagger \alpha}_i c_{i \beta}
- {1\over2} \delta^\alpha_\beta$ are the 15 traceless $SU(4)$ spin generators, 
the analogs of the three Pauli spin matrices in the 
familiar $SU(2)$ case.  With our summation convention
$Tr \{ S(i) S(j) \} \equiv S^{\alpha}_{\beta}(i) S^{\beta}_{\alpha}(j)$.  
The next-nearest-neighbor terms are $O(t^4/U^3)$ and 
$\tan(\theta) = C~ {{t^2}/{U^2}}$, 
where we find the constant $C > 0$; its exact value can be computed\cite{Pert}.
The purely nearest-neighbor $SU(4)$ spin chain was 
studied by Affleck, Arovas, Marston and Rabson\cite{Affleck1}.  
A combination of exact ground states, RG analysis, and conformal field theory
permitted the determination of the entire phase diagram, the antiferromagnetic 
region of which is depicted in Fig. \ref{fig2}.  

\begin{figure}
\begin{multicols}{2}
\epsfxsize = 7.5cm \epsfbox{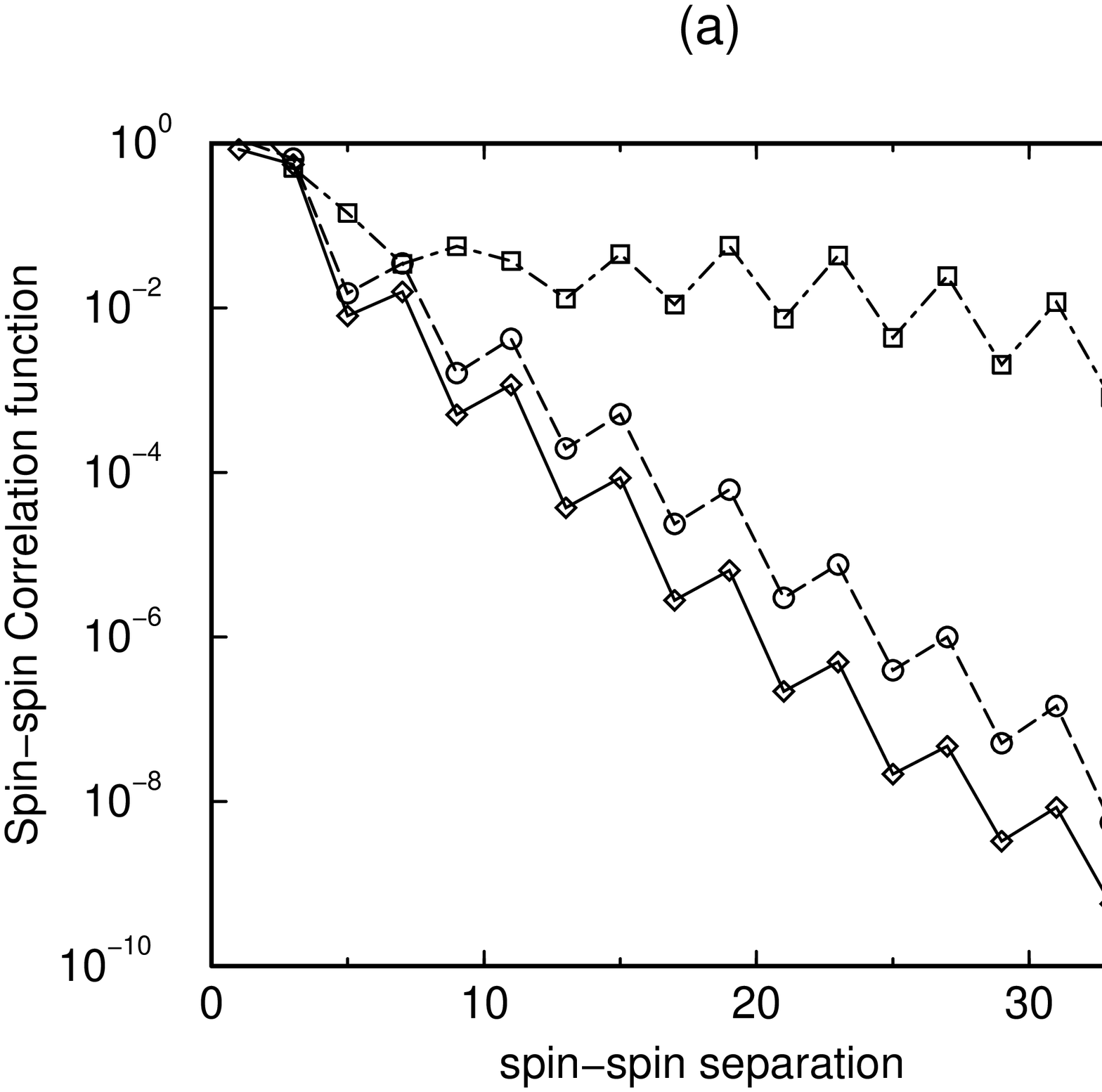}
\epsfxsize = 7.5cm \epsfbox{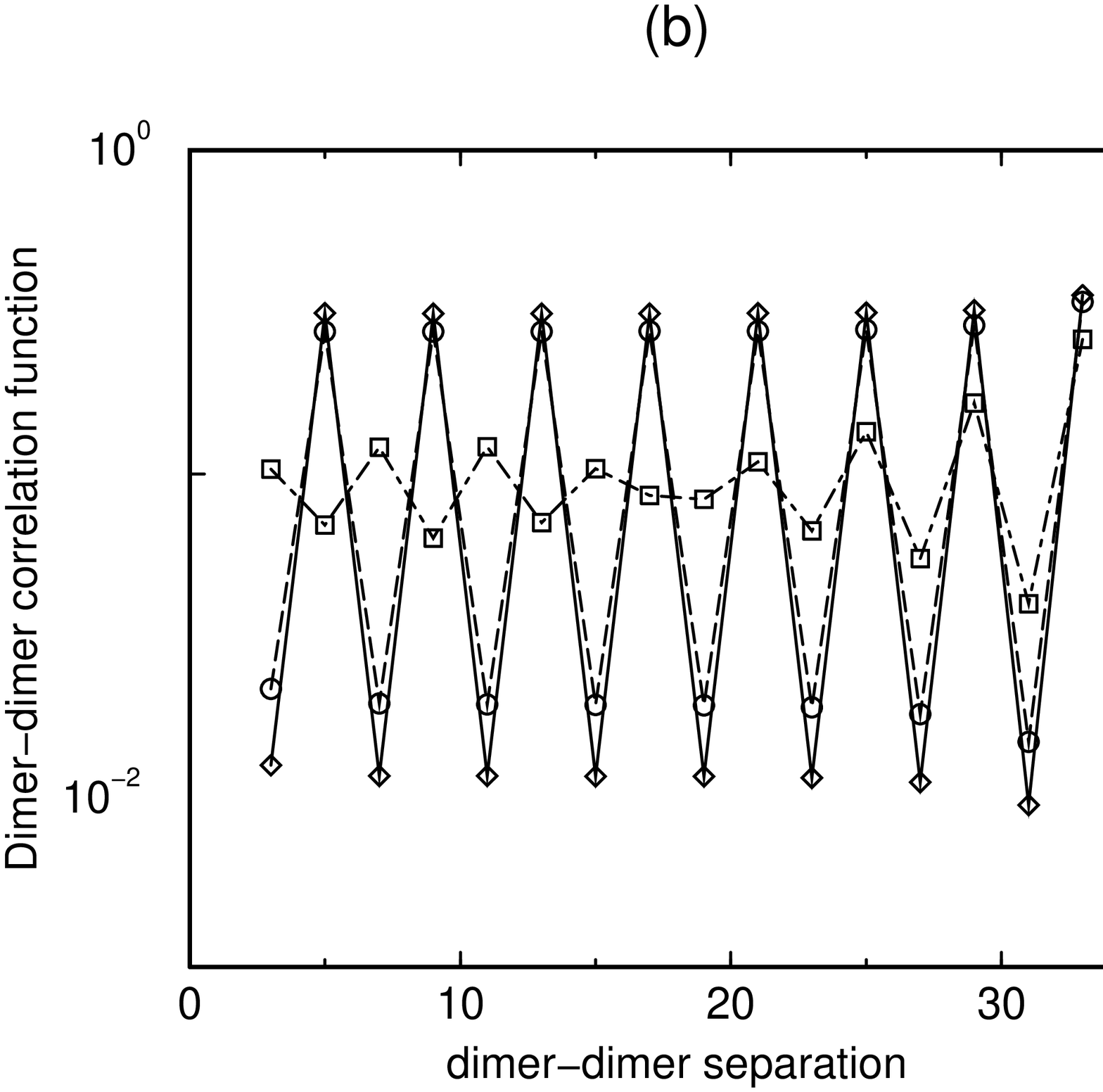}
\end{multicols}
\caption{
DMRG calculation of (a) the spin-spin 
and (b) the dimer-dimer correlation functions for odd lattice separation 
and centered at the middle of a spin chain of length $L = 36$.
Diamonds: $\theta = 0$; circles: $\theta = 0.2$; and squares: $\theta = 0.416$.
The dimerized order diminishes as 
$\theta$ approaches $\theta^* = \tan^{-1}(1/2) \approx 0.4636$}
\label{spdm}
\end{figure}

There is a spin gap at both weak and strong coupling; we therefore expect
the gap to persist at all values of $t/U$. For small values of ${t/U}$, $\theta$
is also small and in the low temperature limit ($T \ll J/k_b \approx 
250^\circ$ K for the $InAs/GaAs$ dot) the system is in a  
dimerized phase\cite{Affleck2} with broken translational symmetry 
which can be qualitatively described
as a set of nearest-neighbor $SU(4)$ singlet bonds as depicted in  
Fig. \ref{fig2}.  Spins not connected by a singlet bond are uncorrelated;
in other words each of the $6 \times 6 = 36$ 
possible configurations of spin and orbital momentum on two such 
sites are realized with equal probability.  In contrast, spins on sites 
connected by a singlet bond are tightly constrained: there is zero
amplitude for the same configuration to be found simultaneously 
on both of the sites.
This has direct experimental consequences as explained below 
at the end of Sec. \ref{section4}.  
The dimerized state, which also breaks reflection symmetry
about site centers, has a large excitation gap since  
$O(J)$ energy is required to break a bond.  
Consequently spin-spin correlations decay exponentially as 
$\langle Tr \{ S(i) S(j) \} \rangle \propto \exp(- |i - j|/{\xi})$, 
where $\xi$ is the spin-spin correlation length. 
White's infinite-size density matrix renormalization Group\cite{White}
(DMRG) analysis with open boundary conditions at the chain ends confirms
this scenario\cite{tobe} and determines $\xi$ to be of order the lattice
spacing at  $\theta = 0$, see Fig. \ref{spdm} (a).  Another quantity of
interest here is dimer-dimer correlation function,
$\langle Tr \{S(i)S(i+1)\} Tr \{S(j) S(j+1)\}\rangle$, which tells us
the probability to find a dimer on the link between sites $i$ 
and $i+1$ given that there is one between sites $j$ and $j+1$.  
The open boundary condition at the chain ends favors one of the two
possible dimerization patterns, see Fig. \ref{spdm} (b).  
The amplitude of dimer-dimer correlation, the difference between its maximum
and minimum values, can be used as an order parameter which provides a
quantitative measure of the degree of dimerization. 

It is interesting to note that dimer order can be achieved in 
ordinary translationally-invariant 
$SU(2)$ antiferromagnetic chains only with large
next-nearest-neighbor or biquadratic exchange.  Here, however, the dimerized 
state at $\theta = 0$ is a natural consequence of the enlarged $SU(4)$ symmetry.
For $\theta > \theta^* = \tan^{-1}(1/2) \approx 0.4636$, 
the chain is in a new phase 
of matter -- not realizable for ordinary $SU(2)$ chains -- 
characterized by spontaneously broken charge-conjugation (C-breaking) symmetry,
a spin gap,  
and extended singlet valence bonds\cite{Affleck1}.  The C-breaking state,
unlike the dimerized state, breaks reflection symmetry about the centers of
bonds, see Fig. \ref{fig2}.  We find that the spin-spin correlation length 
increases, and the dimer-dimer order parameter decreases, 
as the system approaches the transition to the C-breaking phase at 
$\theta = {\theta}^{*}$, see Fig. \ref{spdm} (a) and (b).
It may, however, be difficult to reach 
the C-breaking phase in experimental realizations of the system:
as $t/U$ is increased, terms in the effective
Heisenberg model Eq. (\ref{spinHam}) such as the next-nearest-neighbor exchange
$Tr \{ S(i) S(i+2) \}$ become increasingly important.  This term
has a positive coefficient\cite{Pert}, favoring dimerized order\cite{Note2}.

\section{Symmetry Breaking}
\label{section4}

It is important to establish whether or not the phases of the pure
$SU(4)$ invariant system survive in the presence of symmetry breaking 
processes.  We show that the massive dimerized phase 
is robust in realistic experimental situations.  
The major symmetry breaking process is due to   
electron-electron interactions which lift the 6-fold degeneracy of the 
configurations of the two $p$-electrons on each dot.  
The resulting $\Delta U$ in the Hubbard model
induces $SU(4) \rightarrow SU(2)$ symmetry
breaking both in the on-site energies ($6 \rightarrow 3 \oplus 1 \oplus 2$)
and in the Heisenberg exchange term.  Both perturbations can be incorporated 
by perturbing the $SU(4)$ invariant Hamiltonian Eq. (\ref{spinHam}) with a 
two-body interaction terms of general bilinear form, which in the simplest 
case of a translationally-invariant system can be written as:
\begin{equation}
H_{Hund}^\prime = \sum_{i=1}^L~ 
S^{\alpha}_{\beta}(i)~ T^{\beta \mu}_{\alpha \nu}~ S^{\nu}_{\mu}(i)~ 
+~ \sum_{i=1}^{L-1}~
S^{\alpha}_{\beta}(i)~ \tilde{T}^{\beta \mu}_{\alpha \nu}~
S^{\nu}_{\mu}(i+1)\ .
\label{su2exchange}
\end{equation}
$SU(4)$ invariance is recovered by setting 
$T^{\beta \mu}_{\alpha \nu},~ \tilde{T}^{\beta \mu}_{\alpha \nu} 
\propto \delta^\beta_\nu \delta^\mu_\alpha$;
different choices for the tensors then realize all possible bilinear
$SU(4) \rightarrow SU(2)$ symmetry breaking terms. 
Other less important symmetry breaking processes include:

1. Non-vanishing hopping between states in neighboring dots with
different orbital angular momenta
($t^\alpha_\beta \neq t~ \delta^\alpha_\beta$)
also breaks $SU(4) \rightarrow SU(2)$, as again only spin symmetry remains.
The breaking is minimized in the pillar array due to rotational symmetry
about the vertical axis.
 
2. Spin-orbit coupling by itself breaks
$SU(4) \rightarrow SU(2) \otimes SU(2)$.
However, this effect is small in semiconducting quantum dots\cite{Note1}.

3. Non-magnetic impurities can lift the orbital degeneracy of a dot, breaking
$SU(4) \rightarrow SU(2)$, as only spin symmetry remains intact.
Spin-flip processes, induced by magnetic impurities or external magnetic
fields, break $SU(4)$ all the way down to discrete symmetries.  It is essential
to eliminate both magnetic and non-magnetic
impurities in and around the semiconducting dots.

\begin{figure}
\epsfxsize = 7.5cm \epsfbox{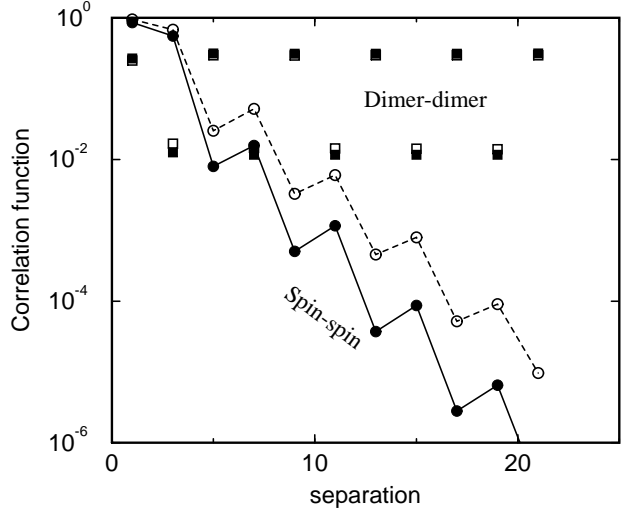}
\caption{
DMRG calculation of the spin-spin (circles)
and the dimer-dimer (squares) correlation functions for odd lattice separation
and centered at the middle of the chain for the case $\theta = 0$ and $L = 36$.
We compare the perfectly $SU(4)$-symmetric chain (filled symbols)
to one with symmetry broken down to $SU(2)$ via $\Delta U \neq 0$
(open symbols).  For the broken symmetry case, the on-site tensor
$T^{\beta \mu}_{\alpha \nu}$ has non-zero entries:
$T^{21}_{12} = T^{12}_{21} = T^{43}_{34} = T^{34}_{43} = J/4$
and
$T^{23}_{14} = T^{41}_{32} = T^{32}_{41} = T^{14}_{23} =
T^{13}_{13} = T^{31}_{31} = T^{24}_{24} = T^{42}_{42} = J/4$.
The nearest-neighbor tensor
$\tilde{T}^{\beta \mu}_{\alpha \nu}$ has non-zero entries:
$\tilde{T}^{33}_{33} = \tilde{T}^{34}_{43} = \tilde{T}^{43}_{34} =
\tilde{T}^{44}_{44} = J/4$
and
$\tilde{T}^{14}_{41} = \tilde{T}^{41}_{14} = \tilde{T}^{23}_{32} =
\tilde{T}^{32}_{23} = J/8$.}
\label{fig3}
\end{figure}

The effects of $SU(4) \rightarrow SU(2)$ symmetry
breaking may be analyzed numerically using the DMRG. 
We find that a block size of $M = 36$
suffices for an accurate description of the massive phases. 
Even for large values of the symmetry breaking parameter
corresponding to $\Delta U \approx J$ the dimer long-range 
order persists, as is evident in Fig. \ref{fig3}. 

Other symmetry breaking mechanisms not included in the general bilinear 
Hamiltonian Eq.(\ref{su2exchange}) can be incorporated by adding a one-body
perturbation to the Hamiltonian Eq. (\ref{spinHam}):
\begin{equation}
\label{one-body}
H^\prime = \lambda^{\beta}_{\alpha}~ \sum_{i=1}^L~  S^{\alpha}_{\beta}(i).
\end{equation}
In particular, spin-orbit coupling corresponds to:
\begin{equation}
\label{spin-orbit}
H^\prime_{SO} = \lambda \sum_{i=1}^L 
[S^1_1(i) - S^2_2(i) - S^3_3(i) + S^4_4(i)]\ .
\end{equation}
We have examined the
effect of this  coupling, Eq. (\ref{spin-orbit}), and have found that even for
an unrealistically large value of $\lambda = J$ the dimerization pattern
remains intact.  The fact that the dimerized phase is robust is not
surprising as the first excited state is separated by a large, of $O(J)$,
energy gap from the ground state. 
 
\begin{figure}
\epsfxsize = 1.5cm \epsfbox{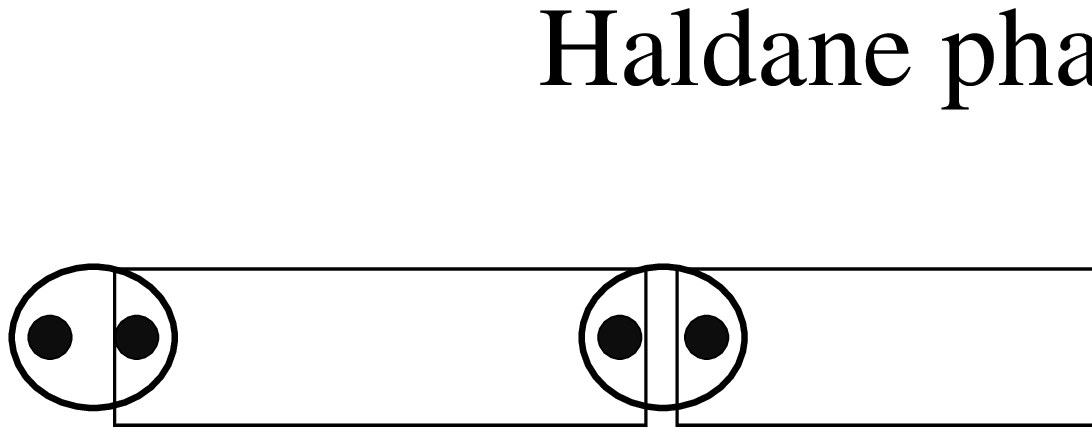}
\caption{
The Haldane gap phase occurs when the $SU(4)$ symmetry is broken
down to the usual $SU(2)$ spin symmetry of a spin-1 quantum antiferromagnet.
$SU(2)$ singlet bonds, depicted as rectangles, involve just two electrons.
Translational symmetry in the ground state is restored, and there are 
free spins at the chain ends both for odd and for even chain lengths.}
\label{haldane-phase}
\end{figure}

In the extreme limit $\Delta U \gg J$ 
of large $SU(4)$ breaking according to Hund's rules,
however, only the triply degenerate $S=1$ states survive and the chain 
is described by an ordinary
spin-1 $SU(2)$ quantum antiferromagnet, which is 
in a different massive phase, the Haldane gap phase\cite{Haldane}, 
with translational symmetry restored as shown in Fig. \ref{haldane-phase}.

Transport measurements can be used to confirm the formation of a Mott-Hubbard
charge gap at half-filling\cite{Newns,DasSarma2}.  To detect the  
dimerized spin structure experimentally, it may be possible to exploit
the fact that, for an odd number of dots only, there are nearly
free spins at the chain ends which will dominate the 
magnetic susceptibility.  This feature distinguishes the dimerized 
state from other possible states of the $SU(4)$ spin chain such as the 
C-breaking and Haldane gap phases which have free spins 
at chain ends for any number of sites [see Fig. \ref{fig2}]. 
The free spins may be observable in sensitive electron-spin resonance 
(ESR) measurements\cite{Hagiwara}, by scanning-tunneling microscopy (STM)
with a magnetized tip,  
or indirectly via optical spectroscopic experiments\cite{Leo}.

\section{Conclusions}
\label{conclusion}

Isolated circular semiconducting dots, filled with a few electrons and 
free of impurities, have already been constructed and
studied\cite{Kouwenhoven}.  We propose the construction of a pillar array of 
such circular dots.  Approximate $SU(4)$ symmetry will be realized if 
some simple requirements are met.  In particular, the dots must be small
to minimize the symmetry breaking effect of the intradot electron-electron
interaction which partially lifts the degeneracy of the 6 different electronic
configurations.  
We predict that a chain of dots, at half-filling (four electrons per dot),
will be in an insulating,
dimerized phase.  Four to six dots will suffice because the correlation length 
is of order the lattice spacing, and  
the state should be robust to various types of symmetry breaking processes 
as there is a non-zero spin gap to low-lying excitations. 
As a practical application of the proposed quantum dot array there is  
the problem of quantum computation\cite{DiVincenzo} which requires a 
high degree of quantum coherence between computing elements.  
The dimerized phase is a strongly 
correlated state and could be used to test the degree of coherence
in an array of quantum dots.  In this it differs greatly from the standard
Coulomb blockade seen in coupled dots which operates independently of 
quantum coherence and, apart from the quantization of the electron charge,
is classical.   
Indeed, quantum many-body phenomena such as the formation of long-range
order are ideal tools to discern quantum coherence.  

Finally we note two possible extensions of this work.  
Experimental evidence for the Kondo effect has been reported in transport
measurements through a single quantum dot\cite{Kondo}.  It would be
interesting to repeat the experiments with a $SU(4)$ dot as increasing the
spin degeneracy enhances the Kondo effect\cite{Hewson,Merino}.  Also,
the III-V dots 
discussed in this paper possess enlarged $SU(4)$ symmetry because the
$2$-fold orbital degeneracy combines with the usual 2-fold spin degeneracy.  
Alternatively, the natural valley degeneracy of silicon\cite{Kittel} could
be exploited.  In $Si$ quantum wells, the 6-fold valley degeneracy is broken
to 2-fold degeneracy by the $Si/SiO_2$ interface\cite{Koester}.  This
remaining degeneracy, like the orbital degeneracy in the III-V dots, is
enough to realize overall $SU(4)$ symmetry.

\acknowledgements 

The authors would like to thank Antal Jevicki,  
Jan\'e Kondev, Sean Ling, and Alex Zaslavsky for fruitful discussions. The 
authors also thank Natalia Onufrieva for help with Fig. \ref{fig1} (A).
Computational work was performed at the Theoretical
Physics Computing Facility at Brown University.  This work was supported in
part by the National Science Foundation through Grants Nos. DMR-9313856 and
DMR-9357613 and by a grant from the Alfred P. Sloan Foundation (J. B. M.).

\end{document}